# A convenient trick to compute cluster sizes in a network

Hsun-Yi Hsieh[1, 2, 3] and Yu-Chun Kao[4]

[1] Great Lakes Bioenergy Research Center, Michigan State University
[2] Kellogg Biological Station Long-Term Ecological Research, Michigan State University
[3] Kellogg Biological Station Long-Term Agroecosystem Research, Michigan State University
[4] U.S. Fish and Wildlife Service

We present a convenient trick for computing the sizes of clusters within a network. The rationale relies on the mathematics of the geometric series and the fundamental matrix of a Markov Chain.

Consider:

$S$ is a $k \times k$ square matrix of direct links between $k$ nodes in a network. $S[i,j] = 0$ suggests no direct link between nodes $i$ and $j$, and $S[i,j] = 1$ suggests otherwise. Then $S^0$, the $k \times k$ identity matrix, represents a matrix of nodes (*i.e.* no direct or indirect connections with other nodes), $S^1$ is the matrix of direct links between node $i$ (row) and node $j$ (column), $S^2$ is the matrix of nodes connected with two degrees of separation…and $S^n$ is the matrix of nodes connected with $n$ degrees of separation. Let *X* denote the sum of matrices $S^0$, $S^1$, $S^2$…$S^n$

$$X = S^0 + S^1 + S^2 + \cdots + S^n \quad (1)$$

Then the count of the non-zero values in row *i* of *X* is the size (*i.e.*, the number of nodes) of the cluster that includes node *i* within *n* degrees of separation. This means that there is no direct or indirect connection between node *i* and node *j* within *n* degrees of separation only when $X[i,j] = 0$.

In practice, calculating X when $n \to \infty$ can be difficult,

Extending the function, we get

$$\sum_{n=0}^{\infty} S^n = S^0 + S^1 + \cdots + S^{\infty} \quad (2)$$

, which denotes the connection in all possible paths of linkage between nodes *i* and *j*.

Solving the geometric series

$$F = \sum_{n=0}^{\infty} S^n = I + S^1 + S^2 + \cdots + S^{\infty} \quad (3)$$

$$F - I = S^1 + S^2 + \cdots = S(S^0 + S^1 + \cdots) = S \times F$$

$$F - SF = I$$

$$F(I - S) = I$$

When *F* converges and is invertible, we obtain

$$F = (I - S)^{-1} \quad (4)$$

This formula shares the format of computing the expected time steps before reaching an absorbing state in a Markov Chain, starting from a transient state.

In the math of the fundamental matrix of a Markov Chain, convergence and invertibility is solved on the basis that a transient matrix is substochastic [1]. We can employ this property to compute the cluster sizes in a network. An easy trick is to transform the matrix by dividing the value of each *S[i,]* by $(i+1)^j$.

$$S[i,] \Rightarrow \frac{S[i,]}{(i+1)^j}$$

The sum of each row in matrix *F* is less than 1 - the only possible exception would be the sum of the first row when the node connects to all of the other nodes. In this case, the maximal sum of this row is equal to the sum of the geometric series

$$\frac{1}{2^1} + \frac{1}{2^2} + \cdots + \frac{1}{2^\infty}$$

, which is equivalent to

$$\frac{1}{2}[\frac{1}{1-\frac{1}{2}}] = 1$$

, which only happens in a network with an unlimited number of nodes. The transformed matrix *S* therefore must be substochastic. We can safely play the math in equation (4). With this, we further obtain the size of each cluster by counting the non-zero elements in row $F[i,]$, in which *i* stands for a node in a network.

## Summary

A Markov process can mathematically relate to network science. This article presents a novel trick applying the math of a Markov Chain to compute the cluster sizes in a network. The described mathematical process advantageously outperforms a brute-force algorithm in computation speed.